\documentclass[letterpaper]{article} 
\usepackage{aaai2026preprint}  
\usepackage{times}  
\usepackage{helvet}  
\usepackage{courier}  
\usepackage[hyphens]{url}  
\usepackage{graphicx} 
\usepackage{natbib}  
\usepackage{caption} 
\usepackage{algorithm}
\usepackage{algorithmic}

\usepackage{newfloat}
\usepackage{listings}
\DeclareCaptionStyle{ruled}{labelfont=normalfont,labelsep=colon,strut=off} 
\floatstyle{ruled}
\newfloat{listing}{tb}{lst}{}
\floatname{listing}{Listing}
\title{A Community-Aware Framework for Influence Maximization \\ with Explicit Accounting for Inter-Community Influence}
\author{
    Eliot W. Robson\textsuperscript{\rm 1},
    Abhishek K. Umrawal\textsuperscript{\rm 2}
}
\affiliations{
    \textsuperscript{\rm 1}Narmi\footnote{Part of this work was completed while Eliot W. Robson was a Ph.D. candidate at the University of Illinois Urbana-Champaign.}\\

    New York, NY 10016 USA\\
    eliot.robson24@gmail.com

    \textsuperscript{\rm 2}University of Illinois Urbana-Champaign\\

    Urbana, Illinois 61801 USA\\
    aumrawal@illinois.edu
}

\usepackage{bibentry}

\usepackage{amsthm}

\newtheorem{problem}{Problem}
\usepackage{multirow}
\usepackage{multicol}
\usepackage{mathtools}

\def\BibTeX{{\rm B\kern-.05em{\sc i\kern-.025em b}\kern-.08em
    T\kern-.1667em\lower.7ex\hbox{E}\kern-.125emX}}

\usepackage[utf8]{inputenc}
\usepackage{pgfplots}
\usepgfplotslibrary{groupplots,dateplot}
\usetikzlibrary{patterns,shapes.arrows}
\pgfplotsset{compat=newest}
\pgfplotsset{scaled y ticks=false}
\usepackage{caption}
\usepackage{subcaption}

\usepackage{amsmath}
\usepackage{mathtools}
\usepackage{amsfonts}
\usepackage{commath}
\usepackage{amsthm}
\usepackage{dsfont}
\usepackage{natbib}
\usepackage{commath}
\usepackage{cleveref}
\usepackage{todonotes}

\newtheorem{theorem}{Theorem}[section]

\newtheorem{lemma}[theorem]{Lemma}
\newtheorem{definition}[theorem]{Definition}
\newtheorem{remark}[theorem]{Remark}

\newcommand{\argmax}{\mathop{\mathrm{arg\,max}}}

\newcommand{\influence}[1]{\operatorname{\sigma}\del{#1}}
\newcommand{\estInfluence}[1]{\operatorname{\widehat{\sigma}}\del{#1}}

\newcommand{\boundedInfluence}[1]{\operatorname{\sigma}^{(2)} \del{#1}}
\newcommand{\estInfluencePart}[2]{\operatorname{\widehat{\sigma}_{#1}}\del{#2}}
\newcommand{\Exp}[2]{\operatorname{\mathbb{E}}_{#1}\sbr{#2}}
\newcommand{\Ind}[2]{Y_{#1}^{#2}}

\newcommand{\CDD}[1]{\operatorname{CDD_{\Partition{}}}\del{#1}}

\newcommand{\TwoNeighbors}[1]{N^{(2)}\del{#1}}
\newcommand{\Comp}[1]{{#1}^c}
\newcommand{\Prob}[1]{\operatorname{\mathbb{P}}\del{#1}}

\newcommand{\Partition}{\mathcal{V}}
\newcommand{\TwoPaths}[2]{\mathcal{P}_{#1,#2}^{(2)}}

\newcommand{\sampleSpaceGraph}{\mathcal{G}}

\DeclareMathOperator{\heuristicFunc}{\rho_{\Partition{}}}
\DeclareMathOperator{\edgeWeightFunc}{\mathcal{E}}
\DeclareMathOperator{\modularity}{Q}

\usepackage{footmisc}
\DefineFNsymbols*{mysymbols}[math]{\dagger\ddagger\S\P\|{\dagger\dagger}{\ddagger\ddagger}}
\setfnsymbol{mysymbols}

\begin{document}

\maketitle

\begin{abstract}
Influence Maximization (IM) seeks to identify a small set of seed nodes in a social network to maximize expected information spread under a diffusion model. While community-based approaches improve scalability by exploiting modular structure, they typically assume independence between communities, overlooking inter-community influence---a limitation that reduces effectiveness in real-world networks. We introduce \text{Community-IM++}, a scalable framework that explicitly models cross-community diffusion through a principled heuristic based on \text{community-based diffusion degree (CDD)} and a \text{progressive budgeting strategy}. The algorithm partitions the network, computes CDD to prioritize bridging nodes, and allocates seeds adaptively across communities using lazy evaluation to minimize redundant computations. Experiments on large real-world social networks under different edge weight models show that Community-IM++ achieves near-greedy influence spread at up to 100 times lower runtime, while outperforming Community-IM and degree heuristics across budgets and structural conditions. These results demonstrate the practicality of Community-IM++ for large-scale applications such as viral marketing, misinformation control, and public health campaigns, where efficiency and cross-community reach are critical.
\end{abstract}

\section{Introduction} \label{section1}

\subsection{Motivation}
The rapid growth of social media has transformed how information, ideas, and products spread across society, influencing domains as diverse as marketing, public health, and civic engagement~\cite{evans2010social}. Organizations increasingly leverage social networks not only for advertising but also for socially beneficial campaigns—such as promoting healthy behaviors, spreading factual information, and raising awareness about critical issues~\cite{goldenberg2001talk, pan2015credit}. A key challenge in these efforts is identifying a small set of individuals whose adoption of a message or product can trigger a large cascade of influence throughout the network.

In real-world scenarios, individuals who act as bridges between communities—often characterized by high betweenness centrality~\cite{freeman1977betweenness}—play a critical role in diffusion. For example, community leaders active in multiple social groups can accelerate vaccination campaigns; fact-checkers who span political communities can curb misinformation during elections; and influencers who engage across diverse interest groups can amplify marketing campaigns beyond niche audiences. Similarly, in disaster response, volunteers connected to multiple local communities can disseminate emergency alerts more effectively. These cases highlight the importance of modeling inter-community influence when designing strategies for information spread.

This challenge is formalized as the Influence Maximization (IM) problem, introduced by \citet{domingos2001mining}: “If we can convince a subset of individuals in a social network to adopt a new product or innovation, and aim to trigger a large cascade of further adoptions, which individuals should we target?” Formally, the goal is to select $k$ seed nodes to maximize the expected number of influenced nodes under a given diffusion model. \citet{kempe2003maximizing} showed that the IM problem is NP-hard, motivating a rich body of research on scalable algorithms. While greedy algorithms offer near-optimal solutions, they rely on costly Monte Carlo simulations, making them impractical for large networks. Heuristic methods improve scalability but often sacrifice accuracy.

To address this trade-off, \citet{uqa-cafsim-23} proposed a community-aware divide-and-conquer framework that partitions the network into communities, optimizes within each, and combines results efficiently. This approach improves runtime while maintaining competitive influence spread. However, a key limitation remains: it overlooks inter-community influence, which can be substantial in real-world networks where information often crosses boundaries, potentially leading to suboptimal strategies.

Our work addresses this gap by introducing a framework that explicitly models inter-community influence, thereby enabling more effective diffusion strategies for applications ranging from viral marketing to public health interventions.

\subsection{Literature Review}
Influence Maximization has been studied extensively under various diffusion models and algorithmic paradigms. Early heuristics such as degree centrality and degree discount~\cite{kkt-msisn-15, chen2009efficient} are computationally efficient but lack theoretical guarantees. Under the \text{independent cascade (IC)} model~\cite{goldenberg2001talk}, \citet{kkt-msisn-15} proposed a greedy algorithm with a $(1-1/e)$-approximation guarantee, later optimized by CELF and CELF++~\cite{leskovec2007cost, gbl-dbasim-11}. Despite these improvements, scalability remains a challenge for large networks.

Community-aware approaches, such as \text{Community-IM}~\cite{umrawal2023leveraging,uqa-cafsim-23}, exploit modular structure to improve efficiency by partitioning the network and applying local optimization. While effective, these methods assume independence between communities, ignoring cross-community influence—a limitation that reduces their applicability in networks with significant inter-community interactions.

Recent work also explores data-driven models~\cite{gbl-dbasim-11, pan2015credit}, fractional budget allocation~\cite{chen2020scalable, umrawal2023fractional,bhimaraju2024fractional}, and online settings~\cite{lei2015online, wen2017online, vaswani2017diffusion,agarwal2022stochastic,nie2022explore}. However, few approaches explicitly address the interplay between community structure and cross-community diffusion, which is central to our contribution.

\subsection{Contribution}
We propose \text{Community-IM++}, an extension of community-aware IM frameworks that incorporates a heuristic for inter-community influence. Our contributions are threefold:
\begin{enumerate}
    \item \textbf{Modeling}: We introduce a principled heuristic to capture cross-community diffusion under the IC model.
    \item \textbf{Algorithmic Framework}: We integrate this heuristic into a scalable divide-and-conquer approach, maintaining efficiency while improving influence spread.
    \item \textbf{Empirical Analysis}: We evaluate Community-IM++ on real-world networks, comparing it against state-of-the-art baselines and analyzing its behavior under different structural and diffusion conditions.
\end{enumerate}

\subsection{Organization}
The remainder of the paper is organized as follows: \Cref{sec:preliminaries} presents preliminaries and problem formulation. \Cref{sec:influence-estimation} introduces our inter-community influence estimation method. \Cref{sec:framework} details the Community-IM++ framework. \Cref{sec:experiments} reports experimental results and insights. Finally, \Cref{sec:conclusions} concludes with future directions.


\section{Preliminaries and Problem Statement}
\label{sec:preliminaries}

In this section, we provide the required preliminaries and formally state the problem of interest.

\subsection{Diffusion Model}
Several models of diffusion over social networks have been proposed in the literature. In this work, we focus on the independent cascade (IC) model~\cite{goldenberg2001talk, gl-ucsa-01}. While other models such as the linear threshold~\cite{granovetter1978threshold, schelling2006micromotives} and pressure threshold~\cite{stutsman2025pressure}, exist, our focus on IC is motivated by the fact that the proposed heuristic is rigorously defined under this model.

In the IC model, we are given a graph $G = (V,E)$ and the random process begins at time $0$ with an initial set $S$ of active nodes, called the seed set. When a node $v \in S$ first becomes active at time $t$, it has a single chance to activate each of its inactive neighbors $w$, succeeding with probability $p_{v,w}$ independently of prior history. If the activation succeeds, $w$ becomes active at time $t+1$. Regardless of the outcome, $v$ cannot attempt to activate $w$ again. The process continues until no new nodes are activated and is progressive, meaning nodes never revert from active to inactive.

\subsection{Influence}
For the IC diffusion process, define the collection of random variables $\{\Ind{S}{(v)}\}_{v \in V}$ where
\[
    \Ind{S}{(v)} =
    \begin{cases}
        1, & \text{if node $v$ is activated with seed set $S$},\\
        0, & \text{otherwise}.
    \end{cases}
\]

The influence $\influence{S}$ of a set $S$ is then defined as
\[
    \influence{S} = \Exp{}{\sum_{u \in V} \Ind{S}{(u)}}.
\]
\citet{kkt-msisn-15} showed that $\influence{S}$ is a monotone non-decreasing submodular set function under the IC model.

\subsection{Problem Statement}
\begin{problem} \label{problem}
The influence maximization (IM) problem is formally defined as:
\begin{align*}
    & \argmax_{S \subseteq V} \influence{S},\\
    \text{s.t.} \quad & |S| \leq k. \qquad \text{(budget constraint)}
\end{align*}
\end{problem}

\section{Estimating the Influence} \label{sec:influence-estimation}
Computing the exact value of $\influence{\cdot}$ is computationally expensive and, in fact, $\#P$-hard. In this section, we define the estimator used in our framework and introduce the concept of diffusion degree to account for inter-community influence.

\subsection{Influence Estimation}
At any time $t$, a node $v \in V$ can be either active or inactive. We denote the process with the random variables $\{\Ind{S}{(v)}\}_{v \in V}$, where $\Ind{S}{(v)}$ is the indicator random variable for $v$ being active at the end of the process with seed set $S$. These random variables are defined over the sample space $\sampleSpaceGraph{}$ of edge activation functions.

The influence $\influence{S}$ is defined as the expected number of active nodes at the end of the cascade, given that $S$ is the seed set:
\[
    \influence{S} = \Exp{\mathcal{G}}{\sum_{v \in V} \Ind{S}{(v)}}.
\]

Since computing $\influence{S}$ exactly is $\#P$-hard~\cite{kkt-msisn-15}, we use the following Monte Carlo estimator:
\[
    \estInfluence{S, k} = \frac{1}{k} \sum_{j=1}^{k} \sum_{v \in V} \Ind{S}{(v)}(g_j),
\]
where $g_1, \dots, g_k \in \sampleSpaceGraph{}$ are sampled uniformly at random. In other words, we approximate influence by averaging the number of activated nodes over $k$ independent simulations.

\subsection{Diffusion Degree}
\label{sec:diffusion_degree}
To explicitly account for inter-community influence—ignored in prior work~\cite{uqa-cafsim-23}—we introduce the concept of diffusion degree.

\begin{definition}
For a node $v \in V$, the diffusion degree, denoted $\boundedInfluence{v}$, is the expected influence of $v$ on nodes within distance two in $G$. Under the independent cascade model:
\[
    \boundedInfluence{v} = \Exp{}{\sum_{u \in \TwoNeighbors{v}} \Ind{\{v\}}{(u)}}.
\]
\end{definition}

\begin{lemma}
For $u, v \in V$, let $\TwoPaths{v}{u}$ be all paths from $v$ to $u$ of length at most two. Then:
\[
    \boundedInfluence{v} = \sum_{u \in \TwoNeighbors{v}} \Big(1 - \prod_{P \in \TwoPaths{v}{u}} (1 - \Prob{P})\Big),
\]
where $\Prob{P}$ denotes the probability that all edges of path $P$ are active.
\end{lemma}

\begin{proof}
For a fixed $u \in \TwoNeighbors{v}$, let $\TwoPaths{v}{u}$ represent the events that each path from $v$ to $u$ is live. Because $u$ is at most two edges away from $v$, these paths are disjoint, and thus the events are independent. Define $I_v$ as the event that $S = \{v\}$. Then:
\begin{align*}
\Prob{u \text{ not activated via } \TwoPaths{v}{u} \mid I_v} &= \Prob{\bigcap_{P \in \TwoPaths{v}{u}} \Comp{P} \mid I_v} \\
&= \prod_{P \in \TwoPaths{v}{u}} (1 - \Prob{P}).
\end{align*}
By linearity of expectation:
\begin{align*}
\boundedInfluence{v} &= \sum_{u \in \TwoNeighbors{v}} \Prob{u \text{ activated via } \TwoPaths{v}{u} \mid I_v} \\
&= \sum_{u \in \TwoNeighbors{v}} \Big(1 - \prod_{P \in \TwoPaths{v}{u}} (1 - \Prob{P})\Big).
\end{align*}
\end{proof}

\begin{remark}
Our formulation of diffusion degree differs from the original definition by \citet{pkm-cmubim-14}, but we retain the name for conceptual similarity. Unlike the original, our definition accounts for multiple independent paths to a node.
\end{remark}

\begin{remark}
Restricting to paths of length at most two is computationally significant: it guarantees path independence, making the heuristic efficient to compute while capturing key inter-community effects.
\end{remark}

\section{The Proposed Framework}
\label{sec:framework}

In this section, we introduce the proposed Community-IM++ framework to solve Problem~\ref{problem}. Our approach builds on the Community-IM framework introduced by \citet{uqa-cafsim-23}, but explicitly accounts for inter-community influence using the machinery developed in \Cref{sec:influence-estimation}. The key contribution is the integration of a heuristic estimator that prioritizes nodes likely to spread influence across community boundaries, addressing a critical limitation of prior work.

\subsection{Overview} \label{sec:algo-overview}
Given a graph $G = (V,E)$ with edge activation probabilities $\edgeWeightFunc : E \to [0,1]$, our algorithm proceeds in four steps:

\begin{enumerate}
    \item[(1)] Obtain a hard partition $\Partition{}$ of the input graph $G$ and $\edgeWeightFunc$ into disjoint communities.
    
    \item[(2)] From this partition, compute a linear set function $\heuristicFunc : V \to \mathbb{R}$ as a heuristic to account for inter-community influence.

    \item[(3)] Construct the final seed set by lazily querying each community for the node with the highest marginal gain until the total budget is exhausted, accounting for inter-community influence of each node using $\heuristicFunc$.
\end{enumerate}

Community-IM++ differs from Community-IM primarily through Step (2), where we introduce a principled heuristic to capture cross-community diffusion. This addition is motivated by real-world scenarios where bridging nodes—those connecting otherwise disconnected communities—play a disproportionate role in spreading information.

\subsection{Community Structure} \label{sec:community-structure}
We partition $G$ into $\Partition{} = \{V_1, \dots, V_c\}$ such that $V_i \cap V_j = \emptyset$ for $i \neq j$. A common measure of partition quality is the modularity score:
\[
    \modularity(\Partition{}, \gamma) = \sum_{i=1}^{c} \Big(m_c - \gamma \frac{K_i^2}{4m}\Big),
\]
where $m_c$ is the total internal edge weight of $V_i$, $m$ is the total edge weight of $G$, and $K_i$ is the weighted degree sum of nodes in $V_i$. The resolution parameter $\gamma$ controls granularity: $\gamma < 1$ favors larger communities, while $\gamma > 1$ favors smaller ones.

We adopt the Leiden algorithm~\cite{twv-fltl-19}, an efficient modularity-based method known for high-quality partitions and scalability. This choice aligns with prior findings that community-aware approaches improve runtime without sacrificing influence spread~\cite{uqa-cafsim-23}.

\subsection{Heuristic Set Function}
Once $\Partition{}$ is computed, we define an estimator for influence within each community:
\[
    \estInfluencePart{i}{S, k, \Partition{}, \heuristicFunc} = \frac{1}{k} \sum_{j=1}^{k} \sum_{v \in V_i} \Big(\Ind{S \cap V_i}{(v)}(g_j^{(i)}) \cdot (1 + \heuristicFunc(v))\Big),
\]
where $\{g_j^{(i)}\}_j$ are samples from the edge activation distribution restricted to $V_i$. The term $(1 + \heuristicFunc(v))$ adjusts for inter-community influence, ensuring that nodes with higher cross-community potential receive greater weight.

\begin{lemma}
For a partition $\Partition{}$, the estimator $\estInfluencePart{i}{S, k, \Partition{}, \heuristicFunc}$ has expected value $\influence{S \cap V_i}$ for an appropriate choice of $\heuristicFunc$ if no pair of nodes $u,v \in V_i$ reachable from $S \cap V_i$ share a common reachable node outside $V_i$.
\end{lemma}

\begin{proof}
If the independence condition holds, we can assign each node $u \in V \setminus V_i$ to a unique $v \in V_i$ reachable from $S \cap V_i$, compute $\Exp{}{\Ind{\{v\}}{(u)}}$, and incorporate this into $\heuristicFunc(v)$. The claim follows from the linearity of expectation.
\end{proof}

\subsubsection{Community-Based Diffusion Degree.}
We instantiate $\heuristicFunc$ using the community-based diffusion degree (CDD), defined as:
\[
    \CDD{v} = \sum_{u \in \TwoNeighbors{v} \cap \Comp{V_i}} \Exp{}{\Ind{\{v\}}{(u)}},
\]
where $v \in V_i$. Intuitively, $\CDD{v}$ measures the expected activation of nodes outside $v$'s community within two hops. This choice is computationally efficient and captures key inter-community effects, as discussed in \Cref{sec:diffusion_degree}.

\paragraph{Justification.}
Restricting to two hops balances accuracy and scalability: it preserves independence assumptions while prioritizing nodes that bridge communities. Empirically, such nodes often correspond to high-betweenness actors~\cite{freeman1977betweenness}, which play critical roles in real-world diffusion scenarios.

\paragraph{Social Relevance.}

Nodes with high diffusion degree often correspond to individuals who bridge communities, similar to those with high betweenness centrality~\cite{freeman1977betweenness}. In real-world networks, such nodes play a critical role in spreading information across otherwise disconnected groups. For example, community leaders active in multiple social circles can accelerate vaccination campaigns, and influencers who span diverse interest groups can amplify marketing or awareness efforts. By incorporating diffusion degree into influence estimation, our framework prioritizes these bridging nodes, enabling strategies that better reflect real-world diffusion dynamics.

\subsection{Generating Candidate Solutions and Progressive Budgeting}
After computing $\heuristicFunc$, we generate nested solutions for each community and combine them using a progressive budgeting algorithm~\cite{uqa-cafsim-23}. This approach allocates seeds incrementally across communities based on marginal gains, ensuring near-optimal influence spread under budget constraints.

\subsubsection{Implementation Details.} Algorithm~\ref{alg:prog_budget} relies on the observation that marginal gains within each community decrease monotonically. We exploit this property by implementing progressive budgeting with lazy evaluation: each community is queried for the next unselected node with the largest marginal gain only when needed. This approach avoids a substantial number of redundant computations, enabling our algorithm to outperform CELF in practice.

Our implementation uses Python coroutines to manage lazy queries efficiently. While this optimization limits parallelism, it reduces overall work performed and memory overhead. CELF~\cite{leskovec2007cost} is used as the subroutine for computing marginal gains within each community.

\begin{algorithm}[h]
\caption{Lazy Progressive-Budgeting}
\label{alg:prog_budget}
\begin{algorithmic}[1]
\REQUIRE $s_1, \dots, s_c, k$
\STATE $\{\delta_i\}_{i=1}^{c} \gets \{\sigma_i (s_i(1))\}_{i=1}^{c}$ \COMMENT{Initialize marginal gains in a heap}
\STATE $\{k_i\}_{i=1}^{c} \gets \{0\}_{i=1}^{c}$ \COMMENT{Initialize the budget allocations}
\STATE $S^* \gets \emptyset$ \COMMENT{Initialize final set}
\FOR{$\ell = 1$ \TO $k$}
    \STATE $m \gets \arg\max_{i \in \{1, \dots, c\}} \delta_i$ \COMMENT{Community with max gain}
    \STATE $\delta_m \gets \sigma_m(s_m(k_m+1)) - \sigma_m(s_m(k_m))$ \COMMENT{Update marginal gain}
    \STATE $k_m \gets k_m + 1$ \COMMENT{Update budget for community $m$}
\ENDFOR
\STATE $S^* \gets \bigcup_{i=1}^{c} S_{i, k_i}$ \COMMENT{Final seed set}
\RETURN $S^*$
\end{algorithmic}
\end{algorithm}

\section{Experiments}
\label{sec:experiments}

We evaluated the performance of our Community-IM++ framework using real-world social networks. This section describes the datasets and their properties, comparison algorithms, experimental setup, and results with discussion.

\subsection{Network Data and Structural Properties}
The real-world network data was obtained from the Stanford Large Network Dataset Collection~\cite{snapnets}. Downloading and caching were automated using the \texttt{Pooch} library~\cite{pooch-20}. Table~\ref{tab:network-properties} summarizes key structural properties of the networks used in our experiments, including node count, edge count, average degree, and modularity for partitions obtained using the Leiden algorithm~\cite{twv-fltl-19}. These properties are relevant to diffusion dynamics and motivate the use of community-aware approaches.

\begingroup
\renewcommand{\arraystretch}{1.2}
\begin{table*}[h!]
\centering
\caption{Structural properties of the networks used in experiments. Note that the modularity score is a property of the computed partition and not of the original network.}
\label{tab:network-properties}
\begin{tabular}{l r r r r r}
\hline
\textbf{Network} & \textbf{Node Count} & \textbf{Edge Count} & \textbf{Avg Degree} & \textbf{Modularity Score} \\
\hline
Deezer & 28,281 & 92,752 & 6.56 & 0.65 \\ \hline 
DBLP & 317,080 & 1,049,866 & 6.62 & 0.81 \\ \hline 
Amazon & 334,863 & 1,851,744 & 11.06 & 0.91 \\ \hline 
\end{tabular}
\end{table*}
\endgroup



High modularity values indicate strong community structure, motivating the use of community-aware algorithms. Amazon, with a modularity of 0.91, is the most modular network, showing tightly clustered product communities that make cross-community influence challenging. DBLP (0.81) also exhibits strong modularity typical of academic collaboration networks, while Deezer (0.65) reflects substantial community structure in social and music-sharing platforms. Modularity above 0.40 signals meaningful structure, and values of 0.70 or higher are considered strong, underscoring that all three networks present significant community-aware optimization opportunities.

For edge weights, we use the weighted cascade (WC) model~\cite{kkt-msisn-15} and the trivalency (TV) model~\cite{gbl-dbasim-11}. In WC, each in-edge for a node $v$ is set to $1/\text{in-degree}(v)$; in TV, each edge weight is drawn uniformly from $\{0.1, 0.01, 0.001\}$.

\subsection{Algorithms Compared}
We compare Community-IM++ against:
\begin{enumerate}
    \item \textbf{Community-IM}~\cite{uqa-cafsim-23}: A community-aware framework ignoring inter-community influence. Note that our benchmarks for this algorithm
    use the performance optimization described in \Cref{alg:prog_budget} to
    provide a more equitable comparison with Community-IM++.
    \item \textbf{CELF}~\cite{leskovec2007cost}: An optimized greedy algorithm with $(1-1/e)$ approximation guarantees.
    \item \textbf{Degree}~\cite{kkt-msisn-15}: A simple heuristic selecting nodes with highest degree.
\end{enumerate}

All algorithms were implemented in Python using \texttt{CyNetDiff}~\cite{rru-cynetdiff-24} for efficient diffusion simulation. As mentioned in \Cref{sec:framework}, we used the Leiden algorithm~\cite{twv-fltl-19} to detect the communities forming hard partitions of the networks under consideration.


\subsection{Experimental Setup}
Budgets tested: $k \in \{5, 20, 100, 200, 400\}$.  
Influence was estimated as the average number of activated nodes over 10{,}000 Monte Carlo simulations per seed set, with 95\% confidence intervals reported.  
Hardware: 8-core Intel Xeon E5-1660v3 CPU @ 3GHz, 64GB RAM.  
Software: Python 3.12.5, \texttt{NetworkX}~\cite{hsp-networkx-08} for graph storage and conversion.

\subsection{Results}
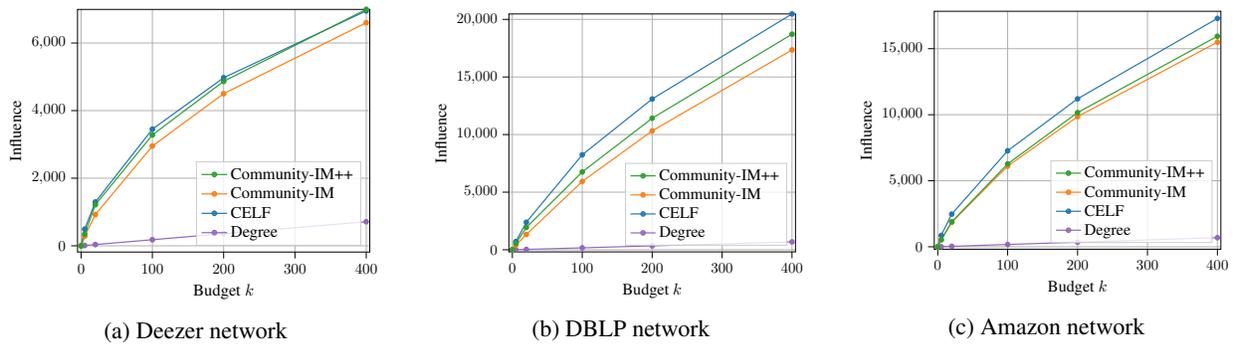
\begin{figure*}[h!]
\centering
\begin{minipage}{.285\textwidth}
  \centering
  \resizebox{\textwidth}{!}{
\begin{tikzpicture}

\definecolor{color0}{rgb}{0.12156862745098,0.466666666666667,0.705882352941177}
\definecolor{color1}{rgb}{1,0.498039215686275,0.0549019607843137}
\definecolor{color2}{rgb}{0.172549019607843,0.627450980392157,0.172549019607843}
\definecolor{color3}{rgb}{0.83921568627451,0.152941176470588,0.156862745098039}
\definecolor{color4}{rgb}{0.580392156862745,0.403921568627451,0.741176470588235}
\definecolor{color5}{rgb}{0.549019607843137,0.337254901960784,0.294117647058824}

\begin{axis}[
legend cell align={left},
reverse legend,
legend style={
  fill opacity=0.8,
  draw opacity=1,
  text opacity=1,
  at={(0.97,0.20)},
  anchor=east,
  draw=white!80!black
},
tick align=outside,
tick pos=left,
x grid style={white!69.0196078431373!black},
xlabel={\large Budget \(\displaystyle k\)},
xmajorgrids,
xmin=-5, xmax=405,
xtick style={color=black},
y grid style={white!69.0196078431373!black},
ylabel={\large Influence},
ymajorgrids,
ymin=-184.6443, ymax=6993.944,
ytick style={color=black}
]

\addlegendentry{\large Degree}
\addplot [thick, color4, mark=*, mark size=1.5, mark options={solid}]
table {%
0 0
5    9.0262   
20   36.8351   
100  179.7345   
200  359.7936   
400  711.3382   
};
\addlegendentry{\large CELF}
\addplot [thick, color0, mark=*, mark size=1.5, mark options={solid}]
table {%
0 0
5    491.6329   
20   1294.8346  
100  3449.3866  
200  4972.7209  
400  6951.0807  
};
\addlegendentry{\large Community-IM}
\addplot [thick, color1, mark=*, mark size=1.5, mark options={solid}]
table {%
0 0
5    280.9443   
20   923.5165   
100  2950.9034  
200  4501.1123  
400  6599.6754  
};
\addlegendentry{\large Community-IM++}
\addplot [thick, color2, mark=*, mark size=1.5, mark options={solid}]
table {%
0 0
5    344.7653   
20   1215.1872  
100  3279.9207  
200  4865.9860  
400  6993.9440  
};
\end{axis}

\end{tikzpicture}}
  \subcaption{Deezer network}
\end{minipage}\hspace{.5cm}
\begin{minipage}{.285\textwidth}
  \centering
  \resizebox{\textwidth}{!}{
\begin{tikzpicture}

\definecolor{color0}{rgb}{0.12156862745098,0.466666666666667,0.705882352941177}
\definecolor{color1}{rgb}{1,0.498039215686275,0.0549019607843137}
\definecolor{color2}{rgb}{0.172549019607843,0.627450980392157,0.172549019607843}
\definecolor{color3}{rgb}{0.83921568627451,0.152941176470588,0.156862745098039}
\definecolor{color4}{rgb}{0.580392156862745,0.403921568627451,0.741176470588235}
\definecolor{color5}{rgb}{0.549019607843137,0.337254901960784,0.294117647058824}

\begin{axis}[
legend cell align={left},
reverse legend,
legend style={
  fill opacity=0.8,
  draw opacity=1,
  text opacity=1,
  at={(0.97,0.20)},
  anchor=east,
  draw=white!80!black
},
tick align=outside,
tick pos=left,
x grid style={white!69.0196078431373!black},
xlabel={\large \large Budget \(\displaystyle k\)},
xmajorgrids,
xmin=-5, xmax=405,
xtick style={color=black},
y grid style={white!69.0196078431373!black},
ylabel={\large Influence},
ymajorgrids,
ymin=-184.6443, ymax=20470.73,
ytick style={color=black}
]

\addlegendentry{\large Degree}
\addplot [thick, color4, mark=*, mark size=1.5, mark options={solid}]
table {%
0    0
5    7.87   
20   32.65   
100  168.33   
200  338.83   
400  683.62   
};
\addlegendentry{\large CELF}
\addplot [thick, color0, mark=*, mark size=1.5, mark options={solid}]
table {%
0    0
5    712.28   
20   2388.27  
100  8243.68  
200  13089.41  
400  20470.73  
};
\addlegendentry{\large Community-IM}
\addplot [thick, color1, mark=*, mark size=1.5, mark options={solid}]
table {%
0    0
5    402.24   
20   1330.75   
100  5930.48  
200  10317.54  
400  17347.32  
};
\addlegendentry{\large Community-IM++}
\addplot [thick, color2, mark=*, mark size=1.5, mark options={solid}]
table {%
0    0
5    546.44   
20   1940.50  
100  6755.21  
200  11416.90  
400  18712.21  
};
\end{axis}

\end{tikzpicture}}
  \subcaption{DBLP network}
\end{minipage}\hspace{.5cm}
\begin{minipage}{.285\textwidth}
  \centering
  \resizebox{\textwidth}{!}{
\begin{tikzpicture}

\definecolor{color0}{rgb}{0.12156862745098,0.466666666666667,0.705882352941177}
\definecolor{color1}{rgb}{1,0.498039215686275,0.0549019607843137}
\definecolor{color2}{rgb}{0.172549019607843,0.627450980392157,0.172549019607843}
\definecolor{color3}{rgb}{0.83921568627451,0.152941176470588,0.156862745098039}
\definecolor{color4}{rgb}{0.580392156862745,0.403921568627451,0.741176470588235}
\definecolor{color5}{rgb}{0.549019607843137,0.337254901960784,0.294117647058824}

\begin{axis}[
legend cell align={left},
reverse legend,
legend style={
  fill opacity=0.8,
  draw opacity=1,
  text opacity=1,
  at={(0.97,0.20)},
  anchor=east,
  draw=white!80!black
},
tick align=outside,
tick pos=left,
x grid style={white!69.0196078431373!black},
xlabel={\large Budget \(\displaystyle k\)},
xmajorgrids,
xmin=-5, xmax=405,
xtick style={color=black},
y grid style={white!69.0196078431373!black},
ylabel={\large Influence},
ymajorgrids,
ymin=-184.6443, ymax=17831.7897,
ytick style={color=black}
]

\addlegendentry{\large Degree}
\addplot [thick, color4, mark=*, mark size=1.5, mark options={solid}]
table {%
0    0
5    8.9862   
20   35.8402   
100  175.1115   
200  347.5821   
400  690.2107   
};
\addlegendentry{\large CELF}
\addplot [thick, color0, mark=*, mark size=1.5, mark options={solid}]
table {%
0    0
5    858.4275   
20   2474.6056  
100  7266.697  
200  11202.728  
400  17290.4123  
};
\addlegendentry{\large Community-IM}
\addplot [thick, color1, mark=*, mark size=1.5, mark options={solid}]
table {%
0    0
5    523.1748   
20   1857.6762   
100  6103.5985  
200  9852.5683  
400  15495.0841  
};
\addlegendentry{\large Community-IM++}
\addplot [thick, color2, mark=*, mark size=1.5, mark options={solid}]
table {%
0    0
5    533.1189   
20   1889.3603  
100  6275.919  
200  10159.0251  
400  15941.5795  
};
\end{axis}

\end{tikzpicture}}
  \subcaption{Amazon network}
\end{minipage}
\caption{Influence vs. Budget ($k$) for different networks under weighted cascade edge-weight model.} \label{fig:ic-wc-influence}
\end{figure*}

\begin{figure*}[h!]
\centering
\begin{minipage}{.285\textwidth}
  \centering
  \resizebox{\textwidth}{!}{
\begin{tikzpicture}

\definecolor{color0}{rgb}{0.12156862745098,0.466666666666667,0.705882352941177}
\definecolor{color1}{rgb}{1,0.498039215686275,0.0549019607843137}
\definecolor{color2}{rgb}{0.172549019607843,0.627450980392157,0.172549019607843}
\definecolor{color3}{rgb}{0.83921568627451,0.152941176470588,0.156862745098039}
\definecolor{color4}{rgb}{0.580392156862745,0.403921568627451,0.741176470588235}
\definecolor{color5}{rgb}{0.549019607843137,0.337254901960784,0.294117647058824}

\begin{axis}[
legend cell align={left},
reverse legend,
legend style={
  fill opacity=0.8,
  draw opacity=1,
  text opacity=1,
  at={(0.97,0.20)},
  anchor=east,
  draw=white!80!black
},
tick align=outside,
tick pos=left,
x grid style={white!69.0196078431373!black},
xlabel={\large Budget \(\displaystyle k\)},
xmajorgrids,
xmin=-5, xmax=405,
xtick style={color=black},
y grid style={white!69.0196078431373!black},
ylabel={\large Influence},
ymajorgrids,
ymin=-18.6443, ymax=1274.0869,
ytick style={color=black}
]
\addlegendentry{\large Degree}
\addplot [thick, color4, mark=*, mark size=1, mark options={solid}]
table {%
0    0
5    5.2763   
20   22.5057   
100  108.4300   
200  218.6594   
400  437.4820   
};
\addlegendentry{\large CELF}
\addplot [thick, color0, mark=*, mark size=1, mark options={solid}]
table {%
0    0
5    81.2605   
20   202.8318  
100  500.9929  
200  744.1892  
400  1112.0078  
};
\addlegendentry{\large Community-IM}
\addplot [thick, color1, mark=*, mark size=1, mark options={solid}]
table {%
0    0
5    78.9931   
20   200.5798   
100  517.8087  
200  782.3672  
400  1165.9453  
};
\addlegendentry{\large Community-IM++}
\addplot [thick, color2, mark=*, mark size=1, mark options={solid}]
table {%
0    0
5    78.1722   
20   212.5347  
100  537.3846  
200  805.1444  
400  1199.3125  
};
\end{axis}

\end{tikzpicture}}
  \subcaption{Deezer network}
\end{minipage}\hspace{.5cm}
\begin{minipage}{.285\textwidth}
  \centering
  \resizebox{\textwidth}{!}{
\begin{tikzpicture}

\definecolor{color0}{rgb}{0.12156862745098,0.466666666666667,0.705882352941177}
\definecolor{color1}{rgb}{1,0.498039215686275,0.0549019607843137}
\definecolor{color2}{rgb}{0.172549019607843,0.627450980392157,0.172549019607843}
\definecolor{color3}{rgb}{0.83921568627451,0.152941176470588,0.156862745098039}
\definecolor{color4}{rgb}{0.580392156862745,0.403921568627451,0.741176470588235}
\definecolor{color5}{rgb}{0.549019607843137,0.337254901960784,0.294117647058824}

\begin{axis}[
legend cell align={left},
reverse legend,
legend style={
  fill opacity=0.8,
  draw opacity=1,
  text opacity=1,
  at={(0.97,0.20)},
  anchor=east,
  draw=white!80!black
},
tick align=outside,
tick pos=left,
x grid style={white!69.0196078431373!black},
xlabel={Budget \(\displaystyle k\)},
xmajorgrids,
xmin=-5, xmax=405,
xtick style={color=black},
y grid style={white!69.0196078431373!black},
ylabel={Influence},
ymajorgrids,
ymin=-184.6443, ymax=6133.43,
ytick style={color=black}
]

\addlegendentry{Degree}
\addplot [thick, color4, mark=*, mark size=1.5, mark options={solid}]
table {%
0    0
5    5.13   
20   29.00   
100  141.53   
200  248.25   
400  468.06   
};
\addlegendentry{CELF}
\addplot [thick, color0, mark=*, mark size=1.5, mark options={solid}]
table {%
0    0
5    2354.16   
20   2342.80  
100  2492.02  
200  2480.63  
400  3390.74  
};
\addlegendentry{Community-IM}
\addplot [thick, color1, mark=*, mark size=1.5, mark options={solid}]
table {%
0    0
5    2700.77   
20   3076.05  
100  4053.42  
200  4817.81  
400  5831.19  
};
\addlegendentry{Community-IM++}
\addplot [thick, color2, mark=*, mark size=1.5, mark options={solid}]
table {%
0    0
5    2611.73   
20   3040.56  
100  4013.12  
200  4800.40  
400  5942.32  
};
\end{axis}

\end{tikzpicture}}
  \subcaption{DBLP network}
\end{minipage}\hspace{.5cm}
\begin{minipage}{.285\textwidth}
  \centering
  \resizebox{\textwidth}{!}{
\begin{tikzpicture}

\definecolor{color0}{rgb}{0.12156862745098,0.466666666666667,0.705882352941177}
\definecolor{color1}{rgb}{1,0.498039215686275,0.0549019607843137}
\definecolor{color2}{rgb}{0.172549019607843,0.627450980392157,0.172549019607843}
\definecolor{color3}{rgb}{0.83921568627451,0.152941176470588,0.156862745098039}
\definecolor{color4}{rgb}{0.580392156862745,0.403921568627451,0.741176470588235}
\definecolor{color5}{rgb}{0.549019607843137,0.337254901960784,0.294117647058824}

\begin{axis}[
legend cell align={left},
reverse legend,
legend style={
  fill opacity=0.8,
  draw opacity=1,
  text opacity=1,
  at={(0.97,0.20)},
  anchor=east,
  draw=white!80!black
},
tick align=outside,
tick pos=left,
x grid style={white!69.0196078431373!black},
xlabel={\large Budget \(\displaystyle k\)},
xmajorgrids,
xmin=-5, xmax=405,
xtick style={color=black},
y grid style={white!69.0196078431373!black},
ylabel={\large Influence},
ymajorgrids,
ymin=-18.6443, ymax=1937.2981,
ytick style={color=black}
]

\addlegendentry{\large Degree}
\addplot [thick, color4, mark=*, mark size=1.5, mark options={solid}]
table {%
0    0
5    5.6206   
20   21.6779   
100  108.3815   
200  216.0283   
400  438.6425   
};
\addlegendentry{\large CELF}
\addplot [thick, color0, mark=*, mark size=1.5, mark options={solid}]
table {%
0    0
5    89.1025   
20   235.8126  
100  688.6596  
200  1112.9157  
400  1781.8100  
};
\addlegendentry{\large Community-IM}
\addplot [thick, color1, mark=*, mark size=1.5, mark options={solid}]
table {%
0    0
5    57.4148   
20   200.3018   
100  655.4850  
200  1100.8046  
400  1826.9494  
};
\addlegendentry{\large Community-IM++}
\addplot [thick, color2, mark=*, mark size=1.5, mark options={solid}]
table {%
0    0
5    52.1215   
20   211.7075  
100  656.1323  
200  1111.6937  
400  1846.8154  
};
\end{axis}

\end{tikzpicture}}
  \subcaption{Amazon network}
\end{minipage}
\caption{Influence vs. Budget ($k$) for different networks trivalency edge-weight model.} \label{fig:ic-tv-influence}
\end{figure*}
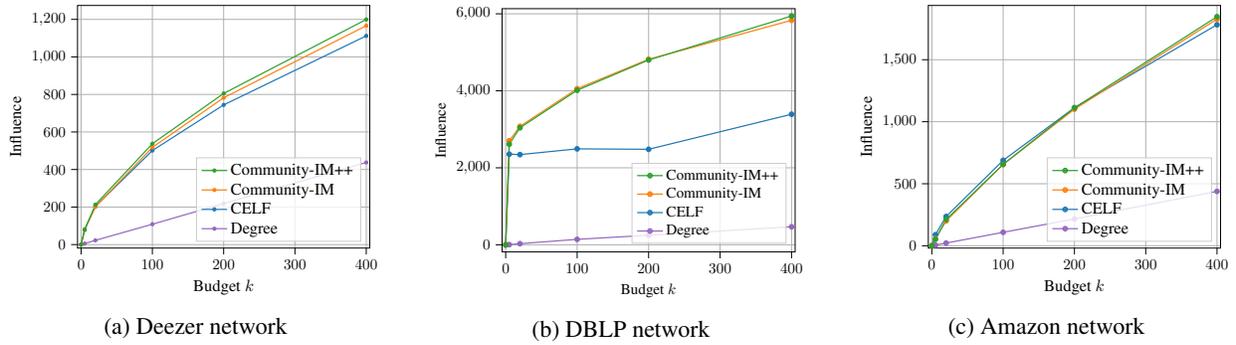

Figures~\ref{fig:ic-wc-influence} and \ref{fig:ic-tv-influence} show influence spread under the WC and TV models, respectively. The $x$-axis represents the seed budget $k$, and the $y$-axis shows the expected number of activated nodes. Each curve corresponds to an algorithm: Degree (purple), CELF (blue), Community-IM (orange), and Community-IM++ (green). Influence was estimated as the average number of activated nodes over 10{,}000 Monte Carlo simulations per seed set, making estimation error negligible.

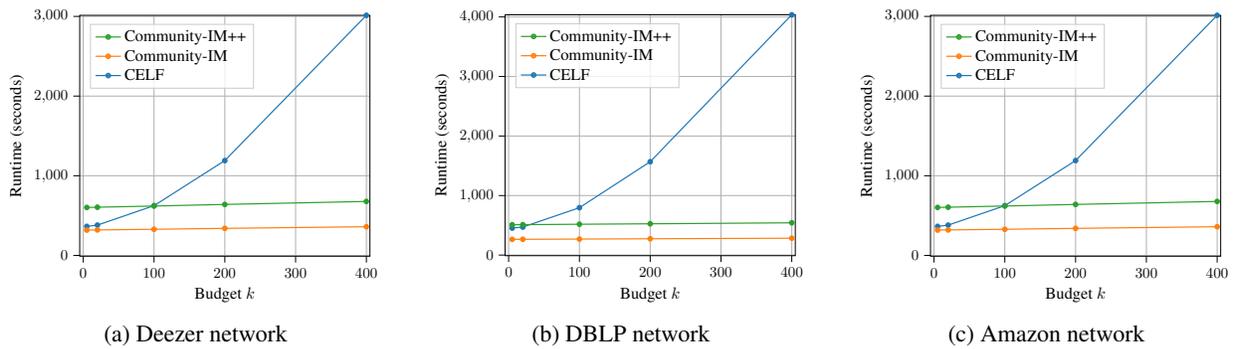
\begin{figure*}[h!] 
\centering
\begin{minipage}{.285\textwidth}
  \centering
  \resizebox{\textwidth}{!}{
\begin{tikzpicture}

\definecolor{color0}{rgb}{0.12156862745098,0.466666666666667,0.705882352941177}
\definecolor{color1}{rgb}{1,0.498039215686275,0.0549019607843137}
\definecolor{color2}{rgb}{0.172549019607843,0.627450980392157,0.172549019607843}

\begin{axis}[
legend cell align={left},
reverse legend,
legend style={
  fill opacity=0.8,
  draw opacity=1,
  text opacity=1,
  at={(0.60,0.83)},
  anchor=east,
  draw=white!80!black
},
tick align=outside,
tick pos=left,
x grid style={white!69.0196078431373!black},
xlabel={\large Budget \(\displaystyle k\)},
xmajorgrids,
xmin=-5, xmax=405,
xtick style={color=black},
y grid style={white!69.0196078431373!black},
ylabel={\large Runtime (seconds)},
ymajorgrids,
ymin=-10, ymax=3013.4415,
ytick style={color=black}
]

\addlegendentry{\large CELF}
\addplot [thick, color0, mark=*, mark size=1.5, mark options={solid}]
table {%
5    364.6362   
20   380.3322  
100  624.5690  
200  1190.4248  
400  3013.4415  
};
\addlegendentry{\large Community-IM}
\addplot [thick, color1, mark=*, mark size=1.5, mark options={solid}]
table {%
5    318.3814   
20   320.1012  
100  328.1018  
200  339.8539  
400  360.1274  
};
\addlegendentry{\large Community-IM++}
\addplot [thick, color2, mark=*, mark size=1.5, mark options={solid}]
table {%
5    602.2460   
20   604.9549  
100  620.2104  
200  640.0416  
400  677.6421  
};
\end{axis}

\end{tikzpicture}}
  \subcaption{Deezer network}
\end{minipage}\hspace{.5cm}
\begin{minipage}{.285\textwidth}
  \centering
  \resizebox{\textwidth}{!}{
\begin{tikzpicture}

\definecolor{color0}{rgb}{0.12156862745098,0.466666666666667,0.705882352941177}
\definecolor{color1}{rgb}{1,0.498039215686275,0.0549019607843137}
\definecolor{color2}{rgb}{0.172549019607843,0.627450980392157,0.172549019607843}

\begin{axis}[
legend cell align={left},
reverse legend,
legend style={
  fill opacity=0.8,
  draw opacity=1,
  text opacity=1,
  at={(0.60,0.83)},
  anchor=east,
  draw=white!80!black
},
tick align=outside,
tick pos=left,
x grid style={white!69.0196078431373!black},
xlabel={\large Budget \(\displaystyle k\)},
xmajorgrids,
xmin=-5, xmax=405,
xtick style={color=black},
y grid style={white!69.0196078431373!black},
ylabel={\large Runtime (seconds)},
ymajorgrids,
ymin=-10, ymax=4036.58,
ytick style={color=black}
]

\addlegendentry{\large CELF}
\addplot [thick, color0, mark=*, mark size=1.5, mark options={solid}]
table {%
5    453.19   
20   470.41  
100  795.82  
200  1568.06  
400  4036.58  
};
\addlegendentry{\large Community-IM}
\addplot [thick, color1, mark=*, mark size=1.5, mark options={solid}]
table {%
5    264.05   
20   265.02  
100  268.87  
200  274.06  
400  283.18  
};
\addlegendentry{\large Community-IM++}
\addplot [thick, color2, mark=*, mark size=1.5, mark options={solid}]
table {%
5    509.59   
20   510.69  
100  518.12  
200  525.51  
400  542.01  
};
\end{axis}

\end{tikzpicture}}
  \subcaption{DBLP network}
\end{minipage}\hspace{.5cm}
\begin{minipage}{.285\textwidth}
  \centering
  \resizebox{\textwidth}{!}{
\begin{tikzpicture}

\definecolor{color0}{rgb}{0.12156862745098,0.466666666666667,0.705882352941177}
\definecolor{color1}{rgb}{1,0.498039215686275,0.0549019607843137}
\definecolor{color2}{rgb}{0.172549019607843,0.627450980392157,0.172549019607843}

\begin{axis}[
legend cell align={left},
reverse legend,
legend style={
  fill opacity=0.8,
  draw opacity=1,
  text opacity=1,
  at={(0.60,0.83)},
  anchor=east,
  draw=white!80!black
},
tick align=outside,
tick pos=left,
x grid style={white!69.0196078431373!black},
xlabel={\large Budget \(\displaystyle k\)},
xmajorgrids,
xmin=-5, xmax=405,
xtick style={color=black},
y grid style={white!69.0196078431373!black},
ylabel={\large Runtime (seconds)},
ymajorgrids,
ymin=-10, ymax=3013.4415,
ytick style={color=black}
]

\addlegendentry{\large CELF}
\addplot [thick, color0, mark=*, mark size=1.5, mark options={solid}]
table {%
5    364.6362   
20   380.3322  
100  624.5690  
200  1190.4248  
400  3013.4415  
};
\addlegendentry{\large Community-IM}
\addplot [thick, color1, mark=*, mark size=1.5, mark options={solid}]
table {%
5    318.3814   
20   320.1012  
100  328.1018  
200  339.8539  
400  360.1274  
};
\addlegendentry{\large Community-IM++}
\addplot [thick, color2, mark=*, mark size=1.5, mark options={solid}]
table {%
5    602.2460   
20   604.9549  
100  620.2104  
200  640.0416  
400  677.6421  
};
\end{axis}

\end{tikzpicture}}
  \subcaption{Amazon network}
\end{minipage}
\caption{Runtime (seconds) vs. Budget ($k$) for different networks under weighted cascade edge-weight model.} \label{fig:ic-wc-runtime}
\end{figure*}

\begin{figure*}[h!] 
\centering
\begin{minipage}{.285\textwidth}
  \centering
  \resizebox{\textwidth}{!}{
\begin{tikzpicture}

\definecolor{color0}{rgb}{0.12156862745098,0.466666666666667,0.705882352941177}
\definecolor{color1}{rgb}{1,0.498039215686275,0.0549019607843137}
\definecolor{color2}{rgb}{0.172549019607843,0.627450980392157,0.172549019607843}

\begin{axis}[
legend cell align={left},
reverse legend,
legend style={
  fill opacity=0.8,
  draw opacity=1,
  text opacity=1,
  at={(0.60,0.83)},
  anchor=east,
  draw=white!80!black
},
tick align=outside,
tick pos=left,
x grid style={white!69.0196078431373!black},
xlabel={\large Budget \(\displaystyle k\)},
xmajorgrids,
xmin=-5, xmax=405,
xtick style={color=black},
y grid style={white!69.0196078431373!black},
ylabel={\large Runtime (seconds)},
ymajorgrids,
ymin=-10, ymax=359.8764,
ytick style={color=black}
]

\addlegendentry{\large CELF}
\addplot [thick, color0, mark=*, mark size=1.5, mark options={solid}]
table {%
5    20.3926   
20   23.3087  
100  61.2437  
200  138.1595  
400  359.8764  
};
\addlegendentry{\large Community-IM}
\addplot [thick, color1, mark=*, mark size=1.5, mark options={solid}]
table {%
5    16.3625   
20   16.9327  
100  19.2130  
200  20.7760  
400  23.9569  
};
\addlegendentry{\large Community-IM++}
\addplot [thick, color2, mark=*, mark size=1.5, mark options={solid}]
table {%
5    31.6555   
20   32.4995  
100  33.9717  
200  37.2310  
400  41.9386  
};
\end{axis}

\end{tikzpicture}}
  \subcaption{Deezer network}
\end{minipage}\hspace{.5cm}
\begin{minipage}{.285\textwidth}
  \centering
  \resizebox{\textwidth}{!}{
\begin{tikzpicture}

\definecolor{color0}{rgb}{0.12156862745098,0.466666666666667,0.705882352941177}
\definecolor{color1}{rgb}{1,0.498039215686275,0.0549019607843137}
\definecolor{color2}{rgb}{0.172549019607843,0.627450980392157,0.172549019607843}

\begin{axis}[
legend cell align={left},
reverse legend,
legend style={
  fill opacity=0.8,
  draw opacity=1,
  text opacity=1,
  at={(0.60,0.53)},
  anchor=east,
  draw=white!80!black
},
tick align=outside,
tick pos=left,
x grid style={white!69.0196078431373!black},
xlabel={\large Budget \(\displaystyle k\)},
xmajorgrids,
xmin=-5, xmax=405,
xtick style={color=black},
y grid style={white!69.0196078431373!black},
ylabel={\large Runtime (seconds)},
ymajorgrids,
ymin=-10, ymax=7948.34,
ytick style={color=black}
]

\addlegendentry{\large CELF}
\addplot [thick, color0, mark=*, mark size=1.5, mark options={solid}]
table {%
5    6104.59   
20   6165.19  
100  6510.05  
200  6942.02  
400  7948.34  
};
\addlegendentry{\large Community-IM}
\addplot [thick, color1, mark=*, mark size=1.5, mark options={solid}]
table {%
5    456.95   
20   458.62  
100  463.32  
200  467.55  
400  473.90  
};
\addlegendentry{\large Community-IM++}
\addplot [thick, color2, mark=*, mark size=1.5, mark options={solid}]
table {%
5    652.62   
20   655.53  
100  659.71  
200  665.65  
400  674.36  
};
\end{axis}

\end{tikzpicture}}
  \subcaption{DBLP network}
\end{minipage}\hspace{.5cm}
\begin{minipage}{.285\textwidth}
  \centering
  \resizebox{\textwidth}{!}{
\begin{tikzpicture}

\definecolor{color0}{rgb}{0.12156862745098,0.466666666666667,0.705882352941177}
\definecolor{color1}{rgb}{1,0.498039215686275,0.0549019607843137}
\definecolor{color2}{rgb}{0.172549019607843,0.627450980392157,0.172549019607843}

\begin{axis}[
legend cell align={left},
reverse legend,
legend style={
  fill opacity=0.8,
  draw opacity=1,
  text opacity=1,
  at={(0.60,0.83)},
  anchor=east,
  draw=white!80!black
},
tick align=outside,
tick pos=left,
x grid style={white!69.0196078431373!black},
xlabel={\large Budget \(\displaystyle k\)},
xmajorgrids,
xmin=-5, xmax=405,
xtick style={color=black},
y grid style={white!69.0196078431373!black},
ylabel={\large Runtime (seconds)},
ymajorgrids,
ymin=-10, ymax=655.1235,
ytick style={color=black}
]

\addlegendentry{\large CELF}
\addplot [thick, color0, mark=*, mark size=1.5, mark options={solid}]
table {%
5    145.9048   
20   149.3754  
100  197.7845  
200  309.0378  
400  655.1235  
};
\addlegendentry{\large Community-IM}
\addplot [thick, color1, mark=*, mark size=1.5, mark options={solid}]
table {%
5    160.5987   
20   161.2970  
100  163.3378  
200  166.2095  
400  171.0610  
};
\addlegendentry{\large Community-IM++}
\addplot [thick, color2, mark=*, mark size=1.5, mark options={solid}]
table {%
5    294.6524   
20   295.3434  
100  298.2158  
200  302.1034  
400  310.0143  
};
\end{axis}

\end{tikzpicture}}
  \subcaption{Amazon network}
\end{minipage}
\caption{Runtime (seconds) vs. Budget ($k$) for different networks trivalency edge-weight model.} \label{fig:ic-tv-runtime}
\end{figure*}
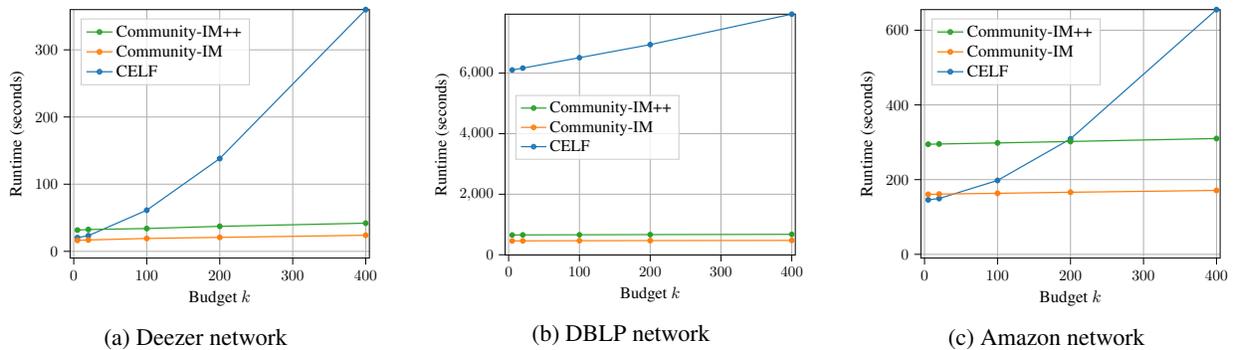

Figures~\ref{fig:ic-wc-runtime} and \ref{fig:ic-tv-runtime} report runtime performance for the same algorithms. The $x$-axis represents the seed budget $k$, and the $y$-axis shows execution time in seconds. 

\subsection{Discussion}
The experimental results reveal several key insights:

\textbf{Influence Spread.} Under WC, Community-IM++ outperforms Degree and Community-IM across all budgets, and approaches CELF’s influence at a fraction of the cost. Under TV, Community-IM++ surpasses all baselines, including CELF, for larger budgets—highlighting the heuristic’s strength in heterogeneous edge-weight settings.

\textbf{Runtime and Scalability.} CELF’s runtime grows rapidly with budget size, becoming impractical for large networks. Community-IM++ exhibits near-constant growth, even for $k=400$, due to progressive budgeting and lazy evaluation. 


\textbf{Cost–Benefit Analysis.} CELF offers marginally higher influence under WC, but at a runtime penalty exceeding 100 times. For practical applications—viral marketing, misinformation control—Community-IM++ provides near-CELF influence at a fraction of the cost.

\section{Conclusion}\label{sec:conclusions}

\textbf{Summary of Contributions.} This work advances community-aware influence maximization by introducing a heuristic that captures inter-community diffusion, addressing a key limitation of prior frameworks. By integrating community-based diffusion degree into a divide-and-conquer approach and coupling it with progressive budgeting, Community-IM++ achieves influence spread comparable to CELF while reducing runtime by orders of magnitude. Our experiments across networks with varying modularity confirm that gains are most pronounced in highly modular structures, where bridging nodes play a pivotal role.

\textbf{Limitations.} While promising, the current implementation is limited to the independent cascade model and a fixed two-hop assumption for inter-community influence. These design choices, while computationally efficient, may restrict generalizability to networks with lower modularity or more complex diffusion dynamics.

\textbf{Future Work.} Future directions include generalization to other diffusion models, such as linear threshold and pressure-based models; ablation and sensitivity analyses of heuristic parameters, including community resolution and hop length; and robustness under alternative community detection methods and overlapping community structures. 

\clearpage
\bibliography{refs}

\end{document}